\documentstyle[preprint,aps]{revtex}
\begin{document}
\draft
%\math-with-secnums
%----------------------------------------------------------------------
\title{ On ``Universal'' Correlations \\
in Disordered and Chaotic Systems}
\author{Jakub Zakrzewski}
\address{
Instytut Fizyki Uniwersytetu  Jagiello\'nskiego, ul. Reymonta 4,
 30-059 Krak\'ow, Poland \\
and Laboratoire Kastler-Brossel, Universit\'e
Pierre et Marie Curie,\\  T12, E1,
4 place Jussieu, 75272 Paris Cedex 05, France
}
\date{\today}
\maketitle
%\receipt{}
%----------------------------------------------------------------------
\begin{abstract}
   Numerical study of the parametric motion of energy levels in
a model system built on Random Matrix Theory is presented. The
correlation function of levels' slopes (the so called velocity
correlation function) is determined numerically and compared
with its limiting analytic form when available. A simple
analytic form of the velocity correlation function is proposed
which very well reproduces numerical data. The results should be
directly applicable in studies of chaotic or mesoscopic
systems.

\end{abstract}
%----------------------------------------------------------------------
\narrowtext

\section{Inroduction}
Random Matrix Theory (RMT) \cite{porter,mehta} can give
accurate predictions concerning
spectral properties of quantum systems chaotic in the classical
limit \cite{fritz,bohigas}, different mesoscopic systems or
metallic disordered particles
\cite{GorEli,Efetov,AltShkl}. The most popular example
is the nearest neighbour spacing distribution, in particular, in its
simplified Wigner form \cite{porter,mehta,fritz,bohigas}.

It is less known that at the beginning of the ``quantum chaos'' area,
the sensitivity of the levels behaviour to an external perturbation
was proposed as a criterion of chaotic behaviour \cite{perci}. The early
numerical studies \cite{pomphrey} considered the distribution of
second
derivatives of the energy levels (or rather their finite difference
approximation) as a candidate for determining a type of
classical dynamics (regular or chaotic) on the basis of quantal spectra.

This idea (and its generalizations) has received renewed and vivid
interest recently. New statistical measures of parametric level
dynamics has been proposed and numerically studied, such as the
distribution of avoided crossings \cite{wilk,gold,ZK91,ZDK93} or the
parametric number
variance \cite{GSBSWZ}. The original second order derivatives of energy
levels, under the name of curvature distribution,
received, arguably, the largest attention following the most
interesting studies of Gaspard and coworkers \cite{grn,grnm}. They
discovered the universality in the parametric behaviour of levels, namely
that the curvature probability density distribution scales
for quantally chaotic systems as
\begin{equation}
P(k)=A_\beta k^{-(\beta+2)}; \quad\beta=1,2,4,
\label{curv}
\end{equation}
in the large $k$ limit.
% An appropriate for a given system
%symmetry universality class (orthogonal, unitary or symplectic) of
%the RMT\cite{porter,mehta,fritz,bohigas} is parameterized by
%$\beta$ in Eq.(\ref{curv}) 
%while
%$k$ is a rescaled dimensionless curvature (see below).
In  Eq.(\ref{curv}), $\beta$ is determined by the symmetry universality
class  (orthogonal, unitary or symplectic) of
the RMT\cite{porter,mehta,fritz,bohigas} appropriate for a given
system while
$k$ is a rescaled dimensionless curvature (see below).
Universality follows since $A_\beta$ is {\it independent} of the
nature of the system or its perturbation. Suppose that energy levels,
$E_i$ of a given system depend on external parameter, $X$.
Define the slopes of the levels (called also velocities in a mechanical
analogy), $V_i(X)=\partial E_i/\partial X$, and their curvatures,
(accelerations) $K_i(X)=\partial^2 E_i/\partial X^2$. Then the
proper scaling is $k_i=K_i\Delta/\langle V^2\rangle$, where $\Delta$
is the mean local level spacing and the average, $\langle ... \rangle$,
is over the different levels and possibly different realizations
of the system. 
  (We assume that the average drift of levels,
    $<V>$, vanishes. One may always redefine level slopes by
substracting the average drift.) 
 Note, for future, that $k$ may be expressed also as
$k_i=\partial^2 e_i/\partial x^2$ where $e_i$ are ``unfolded''
\cite{fritz,bohigas} levels, $e_i=E_i/\Delta$, while $x$ is a
rescaled parameter, $x=X\pi\sqrt{\langle V^2\rangle/\Delta^2}$.

While the large $k$  universal behaviour, (\ref{curv}), was successfully
tested
for a few physical models \cite{grnm,saher,taka1,ZD93},
 it was soon found that
the whole distribution, $P(k)$, is {\it non universal}
\cite{taka1}.
For various chaotic systems small $k$ behaviour of $P(k)$ may be quite
different. The detailed analysis of this interesting behaviour
\cite{ZD93} together with a simultaneous analysis of
avoided crossing distributions \cite{ZDK93} allowed us to link the
nonuniversality
to the partial localization of  wavefunctions on periodic orbits
in chaotic systems (so called scars \cite{hell84}); the
result already alluded to in \cite{taka1}.

One may design other statistical measures of parametric level motion.
Yang and Burgd\"orfer \cite{burg92} introduced in this context the
typical statistical measures of transport such as the diffusion
coefficient or velocity (slopes) autocorrelation function
\begin{equation}
C(X)=\frac{\langle V_i(0)V_i(X)\rangle}{\langle V_i(0)^2\rangle}
\label{vel}
\end{equation}
and argued that in the limit of large $X$ it should behave as $X^{-4}$
(i.e. the diffusion coefficient vanishes in this limit as $X^{-3}$).

A different result has been obtained in a series of papers coming
from Altshuler group \cite{SSA93,SA93,SA932,FSZA93,SLA93,SHCKA}.
Using a supersymmeric approach \cite{Efetov}
they have been able to derive the autocorrelator of density of
states fluctuations (ADS) for all three universality classes
of disordered systems.  They have
found that it takes a universal form, provided the
rescaling discussed above (in terms of $\epsilon_i$ and $x$; note that
our rescaled $x$ differs from that defined in
\cite{SSA93,SA93,SA932,FSZA93,SLA93,SHCKA} by a $\pi$ factor) is
performed. Since the ADS allows, in principle, to
derive various statistical measures, they have formulated a
conjecture   that after the rescaling, statistical
properties of functions of rescaled levels $\epsilon_i(x)$ are universal
(also for strongly chaotic systems)
and independent of the nature of $X$.
This conjecture, although perhaps valid for systems with completely
delocalized wavefunctions is apparently wrong for strongly chaotic
systems \cite{ZDK93,ZD93} where scarred (localized)
 functions are abundant. Such systems show strongly nonuniversal
behaviour.

The nonuniversal features
may bring important information  about the underlying system dependent
structures (scarred wavefunctions, periodic orbits, etc.
\cite{ZDK93,ZD93}). To study them one needs a theory for completely
``random-like'' disordered behaviour. Therefore, results of the MIT
group are of a great value. Unfortunately, the complicated form of
the ADS does not allow to derive analytic results for most of the
interesting statistics except in few limiting cases. For example,
it has been found \cite{SSA93,SA93,SA932} (see also \cite{SzA93})
that velocity autocorrelation
(in rescaled variables) should behave as
\begin{equation}
 C(x)=-\frac{2}{\beta x^2}
\label{limit}
\end{equation}
(recall the $\pi$ difference in the definition of $x$ with respect to
\cite{SSA93,SA93,SA932,FSZA93,SLA93,SHCKA})
in the limit of large $x$, in contradiction with the earlier work
\cite{burg92}. Although theoretical support for (\ref{limit}) has
been also provided within RMT \cite{Ben93}, the numerical data
presented up till now does not allow to draw definitive conclusions
due to poor statistics. 

Importantly, no {\it analytic} expression for $C(x)$ valid in the whole
$x$ domain is known. Simons and Altshuler were able to treat semi-analytically
\cite{SA93,SA932} a closely related quantity, i.e., an autocorrelator
of velocity densities at a specified energy difference: 
\begin{equation}
\label{altshu}
{\cal C}_d(e,x)=\langle{\cal V}_d(e_0,0){\cal V}_d(e_0+e,x).\rangle 
\end{equation}
 Here ${\cal V}_d(e,x)$
is a smoothed (with Gaussian of width $d$) velocity density calculated
at energy $e$ and rescaled parameter $x$ (see \cite{SA932} for details).
 By comparison, $C(x)$ describes the 
correlation of the slope of the level with slope of the {same} level
only at some parameter difference, $x$ while the energy of the
level varies with $x$. For large $x$ one expects both quantities
to coalesce since the net change of $e$ should vanish. This is how
 the limiting behaviour, Eq.(\ref{limit}) may be identified.
Also, in the deep semiclassical limit 
a vanishing difference between the two expressions is expected. 
In principle, however, the two quantities are different. Morover, the
semi-analytical form \cite{SA93,SA932} for ${\cal C}_d(0,x)$ is quite
complicated and given by a triple integral with regularization (which
has to be perfomed numerically), e.g., for
GOE.

It is, therefore, highly desirable to
perform a careful numerical study of velocity distribution for a
generic, ``random-like'' system. Such a  study can be interpretted
as a search for the appropriate distribution by a Monte-Carlo type of
integration. This is the primary aim  of this work. Additionally I
guess a simple analytic form for the resulting distribution.

\section{The numerical approach} 

In the following, I  utilize the  RMT-based model of parametric
dynamics  used before in studies of curvatures, avoided crossings,
and velocity distributions and described in
detail elsewhere \cite{ZDK93,ZD93,fukui}. The model Hamiltonian reads
\begin{equation}
H(X)=H_1 \cos X + H_2 \sin X
\label{ham}
\end{equation}
where matrices $H_i$ are drawn from the appropriate gaussian random
matrix ensemble for each of the $\beta=1,2,4$ cases. Diagonalizations
at different values of $X$ parameter allows me to obtain the velocities
with high accuracy. Since finite size matrices are used, only levels
in the centre of the spectrum are considered to avoid
border effects. Levels are unfolded according to Wigner semicircle
law \cite{porter,mehta,fritz,bohigas}. Several realizations of $H_i$
are used to improve the statistics.
 According to the theory \cite{grnm,ZD93,fukui},
the velocity  distribution is gaussian and the variance
$\langle V(X)^2\rangle$ can be simply estimated analytically. It agrees
with the numerically obtained value to less than half percent.
The variance is independent of $X$, since due to the
form of (\ref{ham}) the averaged density of states is independent of
$X$. Thus velocity correlation function is stationary and does not
depend on the choice of initial $X$ value [the result implicitly
assumed in  Eq.(\ref{vel})  since the left hand side
 does not depend on the
initial choice of the parameter value]. This is important,
especially for large values of the rescaled parameter $x$.
Notice that $H(X=\pi/2)$ is completely decorrelated from $H(0)$
[compare (\ref{ham})] thus the interval of $X$ considered must be
 shorter in length than $\pi/2$. As $\langle V(X)^2\rangle$
grows linearly with $N$ -- the size of matrix of $H(X)$,
 arbitrarily large values of the rescaled, $x$, parameter may
be obtained by increasing $N$. On the other hand,
$H(X=\pi+X_0)=-H(X=X_0)$. That is utilized by calculating
velocities in the whole $[0,\pi]$ interval. Then the statistics is
 enormously increased (especially for small values
of $X$) using the periodicity of the model.

The  converged numerical results are presented in Fig.1, Fig.2,
and Fig.3 for orthogonal ($\beta=1$), unitary ($\beta=2$) and
symplectic ($\beta=4$) ensembles, respectively.
The numerical data for matrices of different sizes tend to  form a
single ``experimental'' curve; it is an indication of the rescaling
property of the purely random model. These data, once tabulated,
 may be used
for comparison with data for physical systems to find and characterize
nonuniversality.

The velocity correlation functions must, obviously, decrease 
for sufficiently small $x$. It is a bit surprizing, however,
 that a definite, and quite large,
negative correlations appear (minima of the curves obtained).
One could imagine that the correlation functions decay smoothly to
zero with increasing $x$. The "correlation hole" deepens and
shifts to smaller values of $x$ with increasing level repulsion
($\beta$). The change of the sign of the level slope (velocity)
happens typically when the level in question is involved in an
avoided crossing (the average slope in the model vanishes). Thus, the
existence of the negative correlation indicates a parameter
(time) scale on which the levels undergo, on average, a single
avoided crossing. This explains the change of the hole position
as a function of $\beta$. One may expect, that for partially
localized systems, the position of the hole may shift to higher
values of the rescaled parameter $x$, since the density of
avoided crossings decreases with increasing order (an abundance
of avoided crossings was proposed a long time ago as an
indicator of quantally chaotic behaviour \cite{noid}).

It would be extremely nice to find a simple analytic formula to
fit the numerical data to an acceptable precision since reproducing
 the data presented in the figures requires a lot of computer time.
Recall the usefulness of simple Wigner formula for the spacing
distributions. The latter may be derived considering the smallest,
$2\times2$ matrices. Our experience with curvature distribution
\cite{ZD93} strongly suggest the nonapplicability of such an
approach for the parametric level motion. Instead, we present below
a simple {\it guess} for the approximate form of the velocity
correlation function. Similarly ``ad hoc'' proposed Brody
distribution \cite{brody} for the spacing distribution in
partially chaotic systems enjoys great popularity and fits the
numerical data very well.

A simplest possible guess which allows to recover the limiting form
of the correlation function, Eq.(3), is a simple algebraic function
with, possibly, some free parameters fitted to the data.
It works quite well for the GOE case \cite{fukui}, however it fails
to describe the ``sharper'' behaviour for other ensembles.
Our proposal comes instead from the ``classical optics'' observation:
the characteristic single change of sign followed by
a minimum (compare the figures) and the $x^{-2}$
 limiting behaviour resembles the
derivative of the dispersion function $d(x)=x/(x^2+a^2$), with
necessary normalization ($a$ - arbitrary). In that case, however, the
minimum is too shallow. On the other hand, in the theory
and practice of spectral
line shapes a quite common function appears, namely the so called plasma
dispersion function \cite{cont}. It describes the convolution of
the gaussian with a complex lorentzian  and is used both
in plasma physics and in atomic spectroscopy (where it appears
once spectral line shapes are averaged over Maxwellian distribution
of atomic velocities). It is a function of a complex variable, $z=x+iy$,
and takes the form
\begin{eqnarray}
Z(z)&=&\pi^{-1/2}\int_{-\infty}^{\infty}dt \frac{\exp(-t^2)}{t-(x+iy)}
 \nonumber\\
&=&i\sqrt{\pi}\exp(-z^2)[1+{\rm erf}(iz)],
\label{Z}
\end{eqnarray}
where erf(.) is a complex error function. Importantly for present
discussion, its imaginary part takes a dispersive ``sharp'' form.
 The depth of the minimum of its complex derivative ${\rm Im}
Z'(z)$
depends on the effective gaussian averaging width, $y^{-1}$ (or
the lorentzian width $y$). I propose,
therefore, that the velocity autocorrelation may be approximated by
\begin{equation}
C(x)=\frac{{\rm Im} Z'(Ax+iy)}{Z'(iy)}
\label{guess}
\end{equation}
where $Z'(z)=-2(1+zZ(z))$ and $Z'(iy)$ [introduced in
denominator of (\ref{guess}) to have $C(0)=1$]   is real for $y>0$.
From the
assumed large $x$ behaviour in the form (\ref{limit}) and known
asymptotic expansions of Z(z) \cite{cont}, one may determine
$A$ in terms of $y$ and $\beta$, and fit $y$ for each $C(x)$ curve.
The dashed line in the figures represent such a fit. The agreement
is very good. In fact it is of similar quality as the Wigner formula
is for the exact RMT spacing distribution.
 The insets in the figures show the large $x$ tails
of $C(x)$ in the logarithmic scale. The large tail behaviour is
in very good agreement with the MIT group prediction
\cite{SSA93,SA93,SA932,Ben93}.

Another motivation for the utilization of the plasma dispersion
function comes from its relation to both lorentzian and gaussian
shapes. The velocity distribution for parametric random ensembles
is gaussian, while the curvature distribution takes a form of a
generalized lorentzian - a Cauchy distribution as guessed
and numerically tested first \cite{ZD93} and then proven 
analytically \cite{vO94,HJS94}.
 A convolution of both these
forms for the velocity correlation function form seems to be quite natural.
 Morover, the fitted values of $y$
for both GUE and GSE are close to $1/\beta$, unfortunately $y$ for
GOE is not close to unity to complete the analogy.

\section{Analytical considerations}

Despite quite good accuracy of the fit, I do not claim that
the proposed expression (\ref{guess}) is an exact velocity autocorrelation
function.
Observe, firstly, that there are some  (small) differences between
the numerical data and the proposed distribution even in the global
scale (compare the figures), especially in the vicinity of the minimum.
Also one may expect, that there are significant qualitative differences
for very small values of the rescaled parameter $x$, although they
are invisible in the figures. In particular, 
for the unitary ensemble, Simons and Altshuler were able to
derive 
a small $x$ expansion  in the form \cite{SA93,SA932}
\begin{equation}
\label{limx}
C_{GUE}(x)\approx 1-2x^2+.... .
\end{equation}
 I used this limiting form  for the unitary ensemble to
determine $y$ and, therefore, fix parameters in \ref{guess} in an unique
way. The agreement
with the numerical data was much worse than for a fit presented in Fig.2.
For that fit, the obtained $x^2$ coefficient is 2.65, distinct from, but 
close to the theoretical value. That indicates that the proposed formula is
approximate only.

While the original derivation of Eq.(\ref{limx}) was performed using
a quite involved supersymmetric calculation, a very simple approach
is possible which also allows one to find the corresponding expression
for the symplectic ensemble, and at the same time, gain some insight
into a different behavior for the orthogonal ensemble. Since the velocity
correlation function is stationary one has 
$<V_i(0)V_i(X)>=<V_i(-X/2)V_i(X/2)>$
where, recall, $X$ is the parameter and $V_i$ the slope of the i-th level
(without unfolding). Assuming the velocity correlation function to be
analytic in the vicinity of $X=0$ one may expand both sides of this
expression up to second power in X. Keeping the same notation as in the
introduction, i.e., $K_i=\partial E_i/\partial X$ for the curvature
and denoting by $K_i^\prime$ the derivative of curvature with respect to $X$
one obtains the following rules for the one-time averages
$<V_i(0)K_i(0)>=0$ and $<V_i(0)K^\prime_i(0)>=-<K_i^2(0)>$ by making
equal the corresponding powers of the small $X$ expansions. Notice that
we obtain that velocites and curvatures are decorrelated at equal times
(in a similar way by considering energies themselves one can verify that 
positions of levels and their velocities are not correlated for random
matrices). Importantly, for the velocity correlation function we obtain 
\begin{equation}
<V_i(0)V_i(X)>=<V^2>-\frac{1}{2}X^2<K_i^2>+... .
\label{new}
\end{equation}
Therefore, the coefficient in front of $X^2$ term is determined by
the variance of the curvature distributions. The distributions are known
exactly for all three ensembles \cite{ZD93,vO94,HJS94}. Calculating 
the  variance and making the unfolding of both energies and the
parameter according to rules given in the introduction one recovers
for GUE the expansion, Eq.(\ref{limx}), while for the symplectic
ensemble one finds
\begin{equation}
C_{GSE}(x)=1-\frac{8}{3} x^2+  ...  .
\label{limse}  
\end{equation}

As far as we know this limiting form appears here for the first time.
Again we may compare the $8/3$ coefficient with the leading term obtained
from the global fit of Eq.(\ref{guess}) which yields 2.29 in even
better agreement than for the corresponding GUE case.

One may notice also that for the orthogonal ensemble the variance
of curvatures diverges \cite{karol}. This shows a non-applicability
of the Taylor expansion around small $X$ for GOE indicating its 
non-analytic behaviour. As it turns out one may prove that $C(x)$
is singular at $x=0$ for all three ensembles, singularities for
GUE and GSE are of higher order in $x$ \cite{my}.
 
\section{Conclusions}

To summarize I have determined numerically the velocity autocorrelation
function for all three universality classes of RMT. The results
may be directly applicable for studies of disordered systems and,
most importantly, for deviations from universality in chaotic systems.
The discrepancy concerning the large $x$ behaviour has been resolved
in favor of the Altshuler and coworkers theory \cite{SSA93,SA93,SA932}.
The analytic formula has been proposed to describe the form of
the velocity correlation function. It may serve as a reference
formula for further studies of this measure since it reproduces
the numerical data very well. Also, as
the complex error
function may be found in almost all software libraries,
its fast and accurate evaluation is very easy.
The formula presented is approximate only as shown discussing
shortly the limiting $x=0$ behaviour of the velocity correlation
function. Finally the limiting analytical form for the symplectic
ensemble, Eq.(\ref{limse}) has been found.

I am  grateful to Dominique Delande for
discussions and patience.
This work was partly supported by the Polish Committee of
Scientific Research under grant P302 102 06.

%----------------------------------------------------------------------
\newpage
%\centerline{\bf References}

%----------------------------------------------------------------------
%\newpage
%\centerline{\bf Figure captions}
\begin{figure}
\caption{
Velocity correlation function for the orthogonal ensemble. Solid line
is a collection of narrowly spaced points for $N=50$, circles and
triangles correspond to $N=200$ and $N=300$, respectively.
Each point corresponds to an average of at least $10^5$ velocities.
Dashed line - fitted formula of Eq.(6), $y=1.317\pm0.005$.
The inset shows the large $x$ tail of the correlation function
in the log-log plot. Stars correspond to $N=50$, other data
as in the main figure.
}
\end{figure}
\begin{figure}
\caption{
Same as Fig.1 but for the unitary ensemble. Fitted parameter
value for the dashed theoretical line is $y=0.50\pm0.005$
}
\end{figure}
\begin{figure}
\caption{
Same as Fig.1 but for the symplectic ensemble. Solid line
(and stars in the inset) correspond to $N=50$ whereas diamonds
represent the data for $N=100$.  Since this
ensemble is most time consuming the procedure of averaging was
stopped when the data showed convergence to save the
computer time. Longer runs would not modify neither the fit
nor the general trend of data. Fitted value of parameter
 is $y=0.25\pm0.008$.
}
\end{figure}
\end{document}